\documentclass[aps,prb,preprint,superscriptaddress]{revtex4}

\usepackage{graphicx}
\usepackage{dcolumn}
\usepackage{bm}
\usepackage[usenames]{color}
\usepackage{amsmath}

\usepackage{amssymb}

\begin{document}

\preprint{preprint}

\title{Duality Breaking of Vortex Configuration in a Hierarchical Honeycomb Network
}

\author{S. Tsuchiya}
\affiliation{Department of Applied Physics, Hokkaido University,
Sapporo 060-8628, Japan}

\author{T. Toshima}
\affiliation{Department of Ecomaterials Engineering, Toyama National College of Technology, Toyama 939-8630, Japan}

\author{H. Nobukane}
\affiliation{Department of Physics, Asahikawa Medical College, Asahikawa, Hokkaido 078-8510, Japan}

\author{K. Inagaki}
\affiliation{Department of Applied Physics, Hokkaido University,
Sapporo 060-8628, Japan}
\affiliation{Center of Education and Research for Topological Science and Technology, Hokkaido University,
Sapporo 060-8628, Japan}

\author{S. Tanda}
\affiliation{Department of Applied Physics, Hokkaido University,
Sapporo 060-8628, Japan}
\affiliation{Center of Education and Research for Topological Science and Technology, Hokkaido University,
Sapporo 060-8628, Japan}

\date{\today}

\begin{abstract}
We report measurements of Little-Parks oscillation on the hierarchical honeycomb-superconducting network for investigating possible effects of hierarchical structure in terms of spatial symmetry, parity and duality. We observed an asymmetric Little-Parks oscillation about $\Phi/\Phi_0 = 1/2$, although spatial symmetry was kept in the network. In comparison with a regular honeycomb network, the asymmetric oscillation is attributed to hierarchy which induces mixture of commensurate and incommensurate regions. The asymmetric oscillation is found to indicate breaking of the duality of vortex configuration. 
\end{abstract}

\maketitle

\section{Introduction}
Complexity, more specifically, hierarchical structure has attracted attention to the problem of decoherence \cite{zurek} leading to the quantum computing \cite{Legg} and the so-called quantum-to-classical transition of cosmological perturbations in the early universe which grew up to galaxies and clusters of galaxies. \cite{astro1,astro2,astro3} Since the hierarchical structure is found in the large-scale structure of galaxies, it may play an important role for the generation of the density fluctuation in the early universe. However, their structural effect on the quantum-to-classical transition have remained unclear.

Superconducting network allows us to perform a proper test for the structural effect of hierarchy. In superconducting networks, the structural properties well affect to the physical properties because this system is sensitive to phase coherence of the order parameter over the network. In fact, phase interference phenomena are driven by the magnetic field known as Little-Parks oscillation. \cite{lpo} Characteristic vortex entry, and configuration of vortices is caused as a result of the structural effect. \cite{squ,one,t3,kk} Therefore one can observe the structural effects as dips or cusps of variation in magnetic field responses. In addition, complex structure can be easily designed.

Generally, in a spatially symmetric system such as regular periodic networks (square, triangular, honeycomb lattices), the duality on these networks is conserved. \cite{hofs,cands} However, in a hierarchical network, it is nontrivial issue whether the duality arises or not. Because their periodicity is different from that of the regular network although spatial symmetry of the hierarchical network is kept. In this letter, we examine possible effects of a hierarchical structure in terms of spatial symmetry, parity and duality of order parameter by using superconducting networks. Our results, which were observed as an asymmetric Little-Parks oscillation about $\Phi/\Phi_0 = 1/2$, indicates breaking of the duality of vortex configuration on the network due to the effect of hierarchical structure.

\section{Experimental}
For our experiment, the so-called hierarchical honeycomb structure, which was discovered in two dimensional charge density wave system was adopted. \cite{toshima} In this structure, the electrons which align on the triangle lattice grow up with hexagonal smoothing clusters hierarchically. The propagation law was applied to superconducting networks. This structure is based on the concept of smoothing and is different from a simple fractal form such as Sielpinski gasket.

\begin{figure}[h]
\begin{center}
\includegraphics[width=0.9\linewidth]{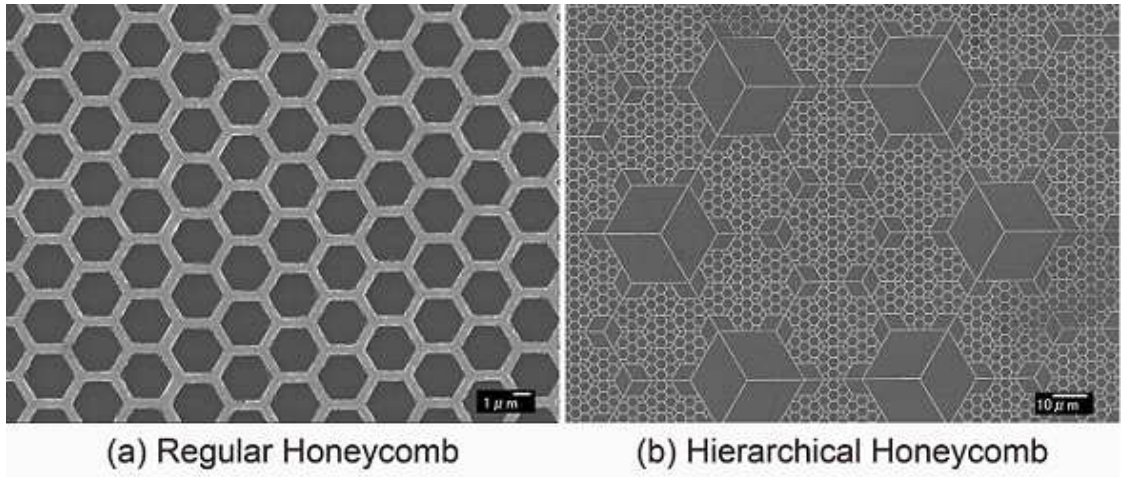}
\caption{SEM image of the samples. (a) The regular honeycomb network, which has about 5000 cells with lattice constant of $2$ $\mu$m, line width of $0.6$ $\mu$m. (b) The hierarchical honeycomb network. The elementary hexagon side length is $2$ $\mu$m with line width of $0.2$ $\mu$m and has $5$ classes of hierarchy.}
\label{fig_1}
\end{center}
\end{figure}

 The lead networks that we used were fabricated by standard electron beam lithography. The gold adhesion layer of $0.01$ $\mu$m and the lead layer of $0.1$ $\mu$m are thermally evaporated on a SiO$_2$ substrate followed by the resist lift-off. To compare hierarchical structure with regular structure, we also prepared a regular honeycomb network in the same way. Fig. \ref{fig_1} shows a scanning electron microscope (SEM) image of the samples. The regular sample has about 5000 cells with lattice constant of $2$ $\mu$m, line width of $0.6$ $\mu$m. In the hierarchical one, the elementary hexagon side length is $2$ $\mu$m with line width of $0.2$ $\mu$m and $5$ classes of hierarchy. 

 Little-Parks oscillation is a powerful tool to investigate the configuration of vortices on the network. Little-Parks oscillation is a periodic variation of superconducting transition temperature ($T_c$) with the magnetic field by the superconducting fluxoid quantization. \cite{lpo} Especially when temperature is near $T_c$, phase coherence is stretched over the whole system. Hence variation of $T_c$ is affected by vortex configuration. Experimentally Little-Parks oscillation of $T_c$ can be observed as a periodic variation of resistance with the magnetic field at fixed temperature, which was taken near the midpoint of normal-to-superconducting transition.

\section{Results and Discussion}

\begin{figure}[h]
\begin{center}
\includegraphics[width=0.5\linewidth]{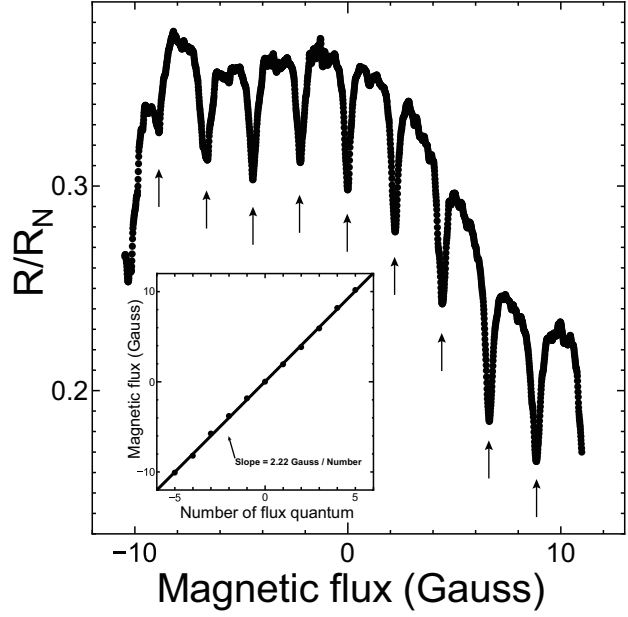}
\caption{The magnetic flux dependence of the sample resistance normalized by the normal state resistance $R_N$. We found periodic dips indicated by the arrows. The inset shows the index number of dip positions as a function of the magnetic flux. The slope shows the period of oscillation as $2.22$ Gauss.}
\label{fig_2}
\end{center}
\end{figure}

 First the regular honeycomb network was investigated as a control experiment. Fig .2 shows the magnetic flux dependence of the sample resistance normalized by the normal state resistance $R_N$. We found periodic dips indicated by the arrows. The inset of fig. \ref{fig_2} shows the index number of dip positions as a function of the magnetic flux. The slope shows the period of oscillation as $2.22$ Gauss. The area calculated from the period is $9.3$ $\mu$m$^2$ and corresponds to a hexagonal unit cell enclosed by the center of the wire. This value compares well to the value $9.6$ $\mu$m$^2$ obtained from SEM observation with $3$ $\%$ accuracy. Thus the period correspond to one-flux quantum $\Phi_0 = h /2e$ per unit cell.

\begin{figure}[h]
\begin{center}
\includegraphics[width=0.5\linewidth]{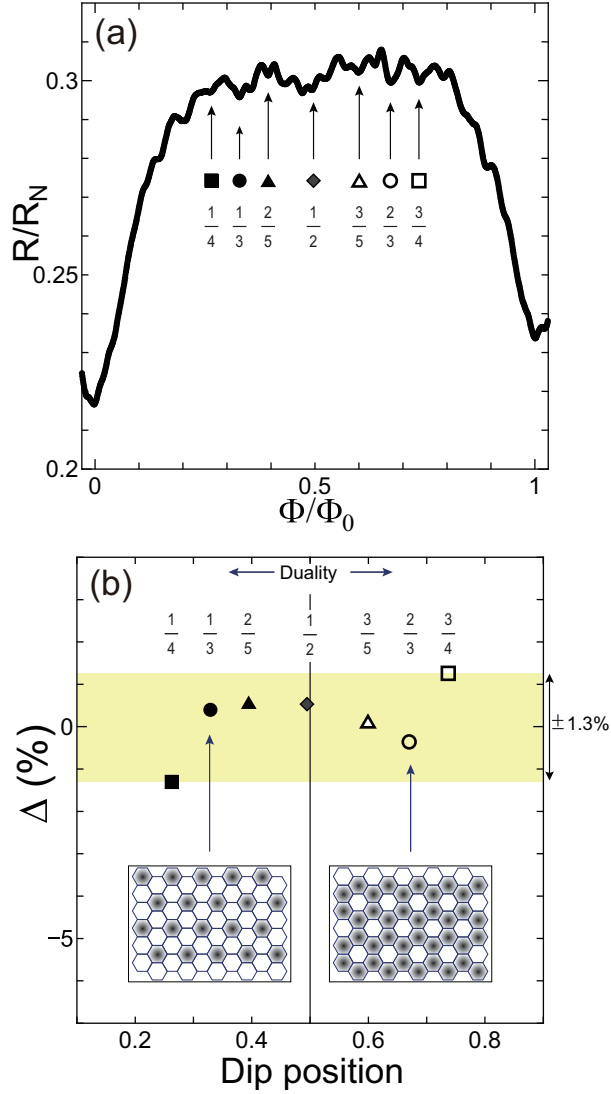}
\caption{(a) The sample resistance as a function of the filling ratio $\Phi/\Phi_0$ in range from $0$ to $1$. The arrows indicates dips with the fundamental filling ratio of $1/4,1/3,2/5,1/2,3/5,2/3$ and $3/4$. (b) Error from Farey sequence versus the dip positions. Shape of symbols denotes correspondence relation in terms of the symmetry about $1/2$. Solid and open symbols denote $0 < \Phi/\Phi_0 < 1/2$ and $1/2 < \Phi/\Phi_0 < 1$, respectively. Solid line in the center is a guide to the eye indicating $\Phi/\Phi_0 = 1/2$. Every dips correspond to $F_5$ within $1.3$ $\%$ accuracy. The inset is vortex configuration on the regular honeycomb network. The unit cells occupied with vortices are shown shaded. Left side is at $\Phi/\Phi_0 = 1/3$ and right side is $\Phi/\Phi_0 = 2/3$. Spatial vortex configuration of $2/3$, is essentially identical to that of $1/3$.}
\label{fig_3}
\end{center}
\end{figure}

 Fig. \ref{fig_3} (a) presents the sample resistance as a function of the filling ratio $\Phi/\Phi_0$, which is the magnetic flux $\Phi$ in units of the flux quantum $\Phi_0$ per a hexagonal unit cell, in range from $0$ to $1$. The arrows indicate dips with the fundamental filling ratio of $1/4,1/3,2/5,1/2,3/5,2/3$ and $3/4$. These fundamental dips appeared in different range, for example, from $1$ to $2$, were not shown in the figure. The results seem like Farey sequence and are consistent with the recent report. \cite{rhc} Fig. \ref{fig_3} (b) exhibits error from Farey sequence $F_5 = \{\frac{1}{5},\frac{1}{4},\frac{1}{3},\frac{2}{5},\frac{1}{2},\frac{3}{5},\frac{2}{3},\frac{3}{4},\frac{4}{5}\}$ versus the dip positions. Shape of symbols denotes correspondence relation from the viewpoint of the symmetry about $1/2$. Solid and open symbols denote $0 < \Phi/\Phi_0 < 1/2$ and $1/2 < \Phi/\Phi_0 < 1$, respectively.  Every dips correspond to $F_5$ within $1.3$ $\%$ accuracy and are clearly symmetric about $\Phi/\Phi_0 = 1/2$. 

\begin{figure}[h]
\begin{center}
\includegraphics[width=0.7\linewidth]{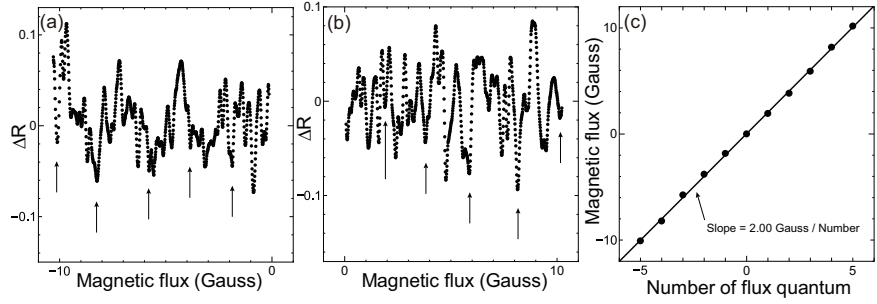}
\caption{The magnetic flux dependence of the sample resistance normalized by $R_N$. (a) In range from $-10$ to $0$ Gauss. (b) In range from $0$ to $10$ Gauss. The arrows indicate periodic dips. (c) The index number of dip positions as a function of the magnetic flux. The slope of the line is  $2.00$ Gauss.}
\label{fig_4}
\end{center}
\end{figure}

 On the other hand, an asymmetric oscillation was observed in the case of the hierarchical honeycomb network. We found a periodic variation of the magneto resistance as shown in Fig. \ref{fig_4} (a) and (b). Measurement noise has been subtracted. The slope of the line is  $2.00$ Gauss as shown in Fig. \ref{fig_4} (c). The area estimated from this period is $10.1$ $\mu$m$^2$ and correspond to just about an elementary hexagon in comparison with $9.8$ $\mu$m$^2$ deduced from SEM observation with $3$ $\%$ accuracy.

\begin{figure}[h]
\begin{center}
\includegraphics[width=0.5\linewidth]{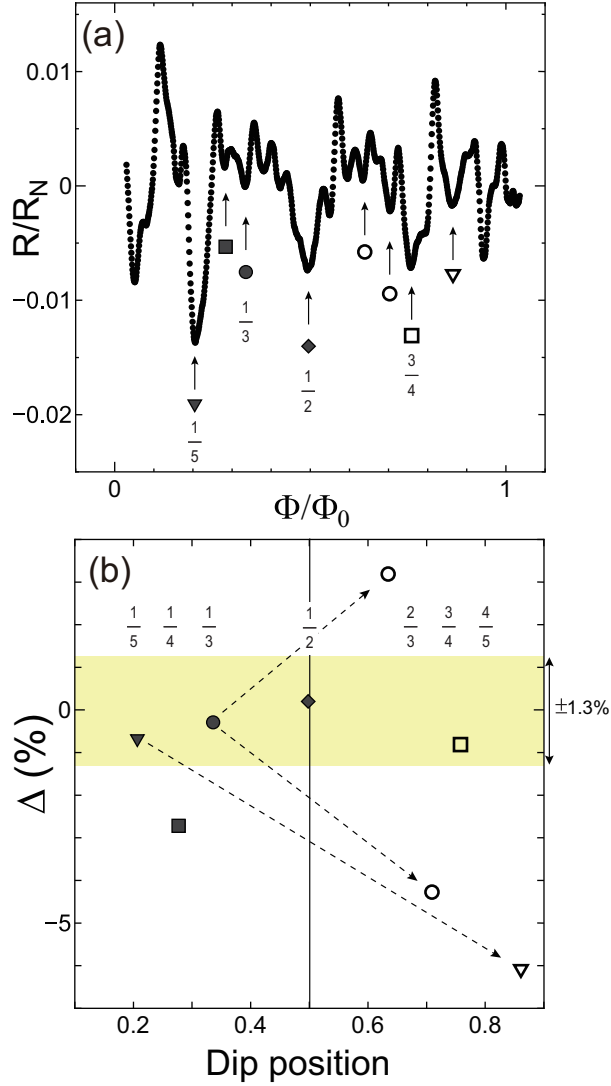}
\caption{(a) The sample resistance as a function of the filling ratio $\Phi/\Phi_0$ in range from $0$ to $1$. The arrows indicate fundamental dips. (b) Error from Farey sequence versus the dip positions. Same notation as in Fig. 3 is used. The dashed arrows are guide to the eye indicating correspondence in terms of the symmetry about $1/2$. Dips of $1/5, 1/3, 1/2$ and $3/4$ correspond to $F_5$ within $1.3$ $\%$ accuracy. The other dips deviate from $F_5$ with accuracy up to $6$ $\%$}
\label{fig_5}
\end{center}
\end{figure}

 Some dips expected from the symmetry about $1/2$ were found to be absent as shown in Fig. \ref{fig_5} (a). In Fig. \ref{fig_5}, measurement noise has been subtracted. Dips indicated by the arrows do not fully agree with Farey sequence. In Fig. \ref{fig_5} (b), the dips of $1/5, 1/3, 1/2$ and $3/4$ correspond to $F_5$ within $1.3$ $\%$ accuracy, the other dips obviously deviate from $F_5$ with accuracy up to $6$ $\%$. This result suggests violation of symmetry about $1/2$ in the case of the hierarchical honeycomb network.

 Now let us discuss the symmetric and asymmetric oscillation about $\Phi/\Phi_0 = 1/2$ by considering vortex configuration on the network. In the case of the regular honeycomb network, vortices are commensurately arranged with its base structure at rational $\Phi/\Phi_0$. For example, at $\Phi/\Phi_0 = 1/3$, one vortex is allocated in every three unit cells which is shown shaded in the inset of Fig. \ref{fig_3} (b). This configuration is strongly pinned and energetically stable. At $\Phi/\Phi_0 = 2/3$, two vortices are allocated in every three unit cells in the same way. When the position occupied with vortex is replaced by the position without vortex, spatial vortex configuration of $2/3$, is essentially identical to that of $1/3$. Both states have energetically same eigenvalue. This is a {\it duality of vortex configuration.} Vortex configuration is globally determined at rational magnetic field. 

\begin{figure}[h]
\begin{center}
\includegraphics[width=0.9\linewidth]{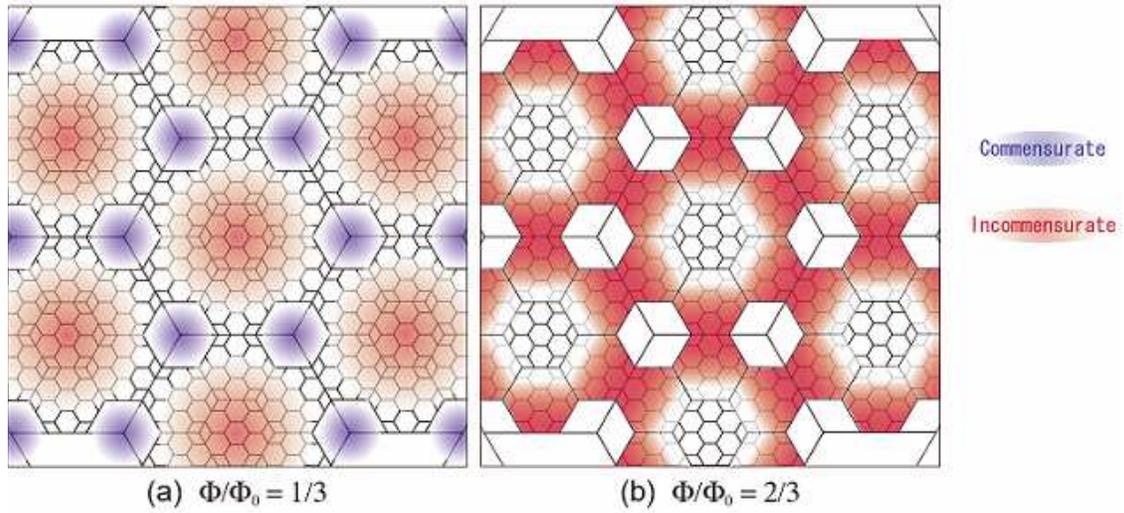}
\caption{The model of vortex configuration on the hierarchical honeycomb network. Blue and red denote commensurate and incommensurate region where vortices are allocated, respectively. Color depth denotes density of vortices. (a) At $\Phi/\Phi_0 = 1/3$. In some regions vortices are commensurately allocated with base structure and other regions are not. (b) At $\Phi/\Phi_0 = 2/3$. In spite of duality operation, the configuration of $2/3$ is not dual for that of $1/3$.}
\label{fig_6}
\end{center}
\end{figure}

 On the other hand, in the case of the hierarchical honeycomb network, the asymmetric oscillation indicates breaking of the duality of vortex configuration although the system has symmetric structure. To explain the asymmetric oscillation we propose a model of vortex configuration as shown in Fig. \ref{fig_6}. Blue and red denote commensurate and incommensurate region where vortices are allocated, respectively. Color depth denotes density of vortices. Fig. \ref{fig_6} (a) shows the configuration at $1/3$, where in some regions vortices are commensurately allocated with base structure and other regions are not, since some different areas exist in the hierarchical honeycomb structure. According to the duality of vortex configuration like the regular case (see the inset of Fig. \ref{fig_3} (b)), the configuration corresponding to $2/3$ is depicted in Fig. \ref{fig_6} (b). In spite of duality operation, the configuration of $2/3$ is not dual for that of $1/3$. There are no commensurate regions because the base structure in incommensurate regions at 1/3 and 2/3 is same. This model could suggest that these states have energetically different eigenvalues and explain the asymmetric oscillation about $1/2$, in particular, absence of dip at $2/3$ in the hierarchical honeycomb network.

 Our result is different from the result of the Sierpinski gasket. \cite{sie} As far as this result, it is found to be symmetric about $1/2$. Therefore, the duality of vortex configuration is conserved. Additionally, vortex configuration is not globally determined at any magnetic field because their characteristic length is absent. Hence the asymmetric pattern is due to the effect of the hierarchical honeycomb structure in itself.

 Finally we have some comments for general properties of our hierarchical structure. Strong pinning of vortices means phase fluctuation of the order parameter is very small or its phase is determined spatially and temporally. If similar hierarchical structure is existed in some systems, the regions where its phase is determined could be appeared spontaneously with spatial dependence. This concept might be applied to the problem of decoherence in complex systems.

\section{Summary}

We measured Little-Parks oscillation in the hierarchical superconducting network for investigating effects of hierarchical structure in terms of spatial symmetry, parity and duality. We observed the asymmetric Little-Parks oscillation about $\Phi/\Phi_0 = 1/2$, although spatial symmetry was kept in the network. In comparison with a regular honeycomb network, the asymmetric oscillation is attributed to hierarchy which induces mixture of commensurate and incommensurate regions. The asymmetric oscillation is found to indicate breaking of the duality of vortex configuration.

\begin{acknowledgements}
The authors are grateful to A. J. Leggett, F. Nori and K. Semba for useful discussions. We also thank K. Yamaya, S. Takayanagi and N. Matsunaga for experimental support. This work was supported Grant-in-Aid for the 21st Century COE program "Topological Science and Technology". 
\end{acknowledgements}

\newpage

\newpage

\newpage

\newpage

\newpage

\newpage

\end{document}